\documentclass[aps,prl,twocolumn,,superscriptaddress,noshowpacs]{revtex4}
\usepackage{hyperref}
\usepackage[utf8]{inputenc}
\usepackage{amsmath}
\usepackage{amsfonts}
\usepackage{amssymb}
\usepackage{graphicx}

\begin{document}
\title{Non-quadratic transverse magnetoresistance of the nodal line Dirac semimetal InBi}
\author{S.V. Zaitsev-Zotov}
\email[]{serzz@cplire.ru} 
\affiliation{Kotelnikov Institute of Radioengineering and Electronics of Russian Academy of Sciences, Mokhovaya 11, bld. 7, Moscow 125009 Russia}

\author{I.A. Cohn}
\affiliation{Kotelnikov Institute of Radioengineering and Electronics of Russian Academy of Sciences, Mokhovaya 11, bld. 7, Moscow 125009 Russia}

\date{\today}

\begin{abstract}
The transverse magnetoresistance of a nodal line Dirac semi-metal InBi has been studied. It is found that the magnetoresistance is not quadratic. In the region of small magnetic fields $ B\lesssim 0.1$~T, it is characterized by high curvature, in the region of medium magnetic fields it is described by the sum of linear and quadratic contributions, and in the region of large magnetic fields $ B\gtrsim 1$~T, it approaches a quadratic law with a curvature several times smaller its zero field value. A phenomenological equation is proposed that allows to describe the entire dependence of the resistance on the magnetic field with an error not exceeding the measurement error of several percent.\end{abstract}

\maketitle

The study of materials with non-trivial topology of the energy structure  is one of the most actively developing fields in solid state physics. Many phenomena considered in the field theory turned out to be accessible for realization and study in topologically nontrivial materials and structures. In particular, the realization of Dirac and Weyl fermions in solids allows one to study in laboratory experiments some exotic states predicted in the field theory. Of great interest are attempts to implement anyons, including Majorana states, as well as the chiral anomaly \cite{review}.

Quasiparticles with the Dirac spectrum arise in a number of materials. Well-known examples are graphene, topological insulators, Dirac semimetals \cite{diracsm}. More recently, it has been found that there are also materials in which the vertices of the Dirac cone are not at one or more points of the Brillouin zone, but form the line \cite{line}. A feature of Dirac node-line semimetals is the much higher density of Dirac states than in materials with Dirac points, which allows us to hope for a more vivid manifestation of the properties due to Dirac fermions. Currently, the existence of such states is proved for a relatively small set of materials, such as PbTaSe$_2 $ \cite{pbtase}, PtSn$_4 $ \cite{ptsn}, ZrSiS, ZrSiTe \cite{Hu, Ali}, as well as InBi \cite{arpes}, the study of the transverse magnetoresistance of which is the subject of this work.

In magnetotransport measurements, many specific features of topological materials are manifested. Much of the effort in this area is directed towards searching and studying the chiral anomaly. The effect develops in parallel electric and magnetic fields, manifests itself in the appearance of negative longitudinal magnetoresistance, and is considered one of the key manifestations of the Weyl fermions in the transport properties \cite{chiral}. As for the transverse magnetoresistance, no such bright effects are expected. Nevertheless, for nodal-line Dirac semi-metals, a number of unusual properties were also predicted and/or discovered experimentally. Among them, an extremely large positive quadratic transverse magnetoresistance exceeding 2 orders of magnitude at helium temperatures in InBi \cite{MR} and 3 orders of magnitude in ZrSiS \cite{Ali}, as well as the presence of a linear contribution to the transverse magnetoresistance in ZrSiSe and ZrSiTe \cite{Ali} were observed.

InBi has a tetragonal unit cell belonging to the non-symmorphic spatial group of symmetry P4/nmm \cite{space_group}. The lattice parameters depend on temperature and vary in the range $a = 4.9589 - 5.0101$ \AA , $ c = 4.8396 - 4.7824$ \AA \ when the temperature changes from 15 to 300 K \cite{ac-T}. First-principle calculations show that InBi is a Dirac semimetal in which the vertices of the Dirac cone form a line in the momentum space along the directions MA and XR of the Brillouin zone, i.e. in the direction along the $c$ axis \cite{arpes}. The presence of the Dirac line in the energy spectrum is confirmed by the results of the ARPES study \cite{arpes}. The study of magnetotransport indicates the presence of an extremely large positive transverse quadratic magnetoresistance, which exceeds 2 orders of magnitude at helium temperatures in the configuration $ B \perp I \parallel ab $ and does not saturate in high magnetic fields \cite{MR}. The absence of  saturation \cite{MR} and its anomalously high value are associated with the equality of the concentrations of electrons and holes in this material, whose mobility at helium temperatures reaches $1.5 \times 10^4 $~cm $^2 $/V $\cdot $ s \cite{MR}.

In this work, we present the results of studying the transverse magnetoresistance in InBi in the configuration $ B \perp I \parallel ab $ with high accuracy. The results obtained made it possible to distinguish features not previously observed in similar measurements in this material. In particular, it was found that the dependence of the resistance on the magnetic field  differs from the quadratic dependence in the entire range of the studied magnetic fields. A phenomenological equation is proposed that describes all the features obtained with a relative error of the order of several percent. The presence of a linear contribution, discovered in this work, corresponds to the expected behavior of the magnetoresistance of Dirac semimetals with a nodal line \cite{LMR}.

\begin{figure} 
\begin{center}
\includegraphics[width=0.45\textwidth]{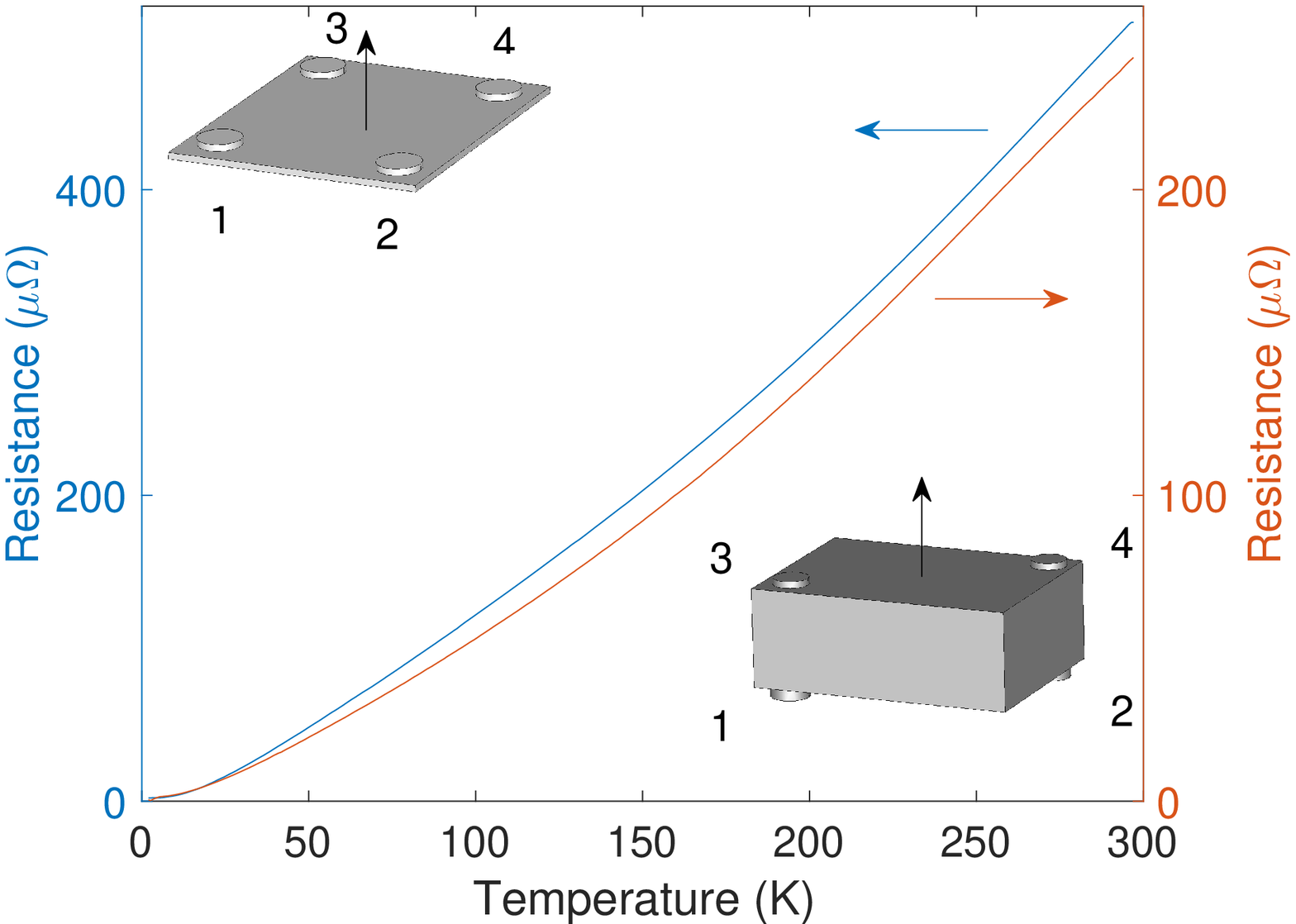}
\caption{Temperature dependences of the resistance of the studied samples. The shape of the samples and the contact positions are shown in the insets. The direction of the $c$ axis and the magnetic field are shown by black arrows. Sample 1 --- the right scale and the lower right corner, sample 2 --- the left scale and the upper left corner. }
\label{rt}
\end{center}
\end{figure}

A stoichiometric mixture of In and Bi (purity 99.998 \% and 99.995 \%, respectively) was placed in a quartz ampoule, evacuated to a pressure of $2\times 10^{- 5}$ Torr, melted for degassing, and sealed off. After thoroughly mixing the melt at a temperature of 200 $^\circ$C, crystallization was carried out in a temperature gradient with the vertical or horizontal position of the ampoule during smooth cooling from 300 to 30 $^\circ$C for 6 hours. As a result, a crystal grew in which the $ c $ axis was tilted to the axis of the ampoule. X-ray diffraction analysis showed the single-phase nature of the grown crystals. Samples for the research were obtained by cleaving in crystallographic directions. The plane of the light cleavage corresponds to the direction (001) \cite{MR,arpes,growth}. To minimize plastic deformation, cleavege was carried out at the temperature of liquid nitrogen. The contacts to the samples were soldered with a low-melting-point solder with melting temperature of 58~$^\circ$C, which was significantly lower than the InBi melting temperature ($\sim 110^\circ$C). In total, 4 samples of different sizes were studied with varying degrees of detail, which showed similar behavior. The results presented in this work were obtained on samples of approximately rectangular shape with sizes of $2.8\times 3.0 \times 1.7$ mm $^3 $ (sample 1) and $ 5 \times 4.5 \times 0.10$ mm $^3$ (sample 2, in both cases, the last size indicated corresponds to the $ c $ axis direction). The location of the contacts on the samples is shown in the insets to Fig. \ref{rt}. The resistance measurements in sample 1 (the lower right corner in the figure \ref{rt}) were carried out by passing the current through contacts 1 and 2 located on one face of the sample, and measuring the voltage between contacts 3 and 4 located on another one. In the case of sample 2 (the upper left inset in the figure \ref{rt}), the current was passed through contacts 1 and 2 located near one of the long edges of the sample, and the voltage was measured between the contacts 3 and 4 located near another long edge.

Fig. \ref{rt} shows temperature dependences of the resistance of the crystals under study. The dependencies are typical for metals and close to the previously observed ones \cite{MR,lal}. The ratios $RRR=R(300{\rm \ K})/R (4.2 {\rm \ K}) $ for the studied samples are 230 and 250.

\begin{figure} 
\begin{center}
\includegraphics[width=0.45\textwidth]{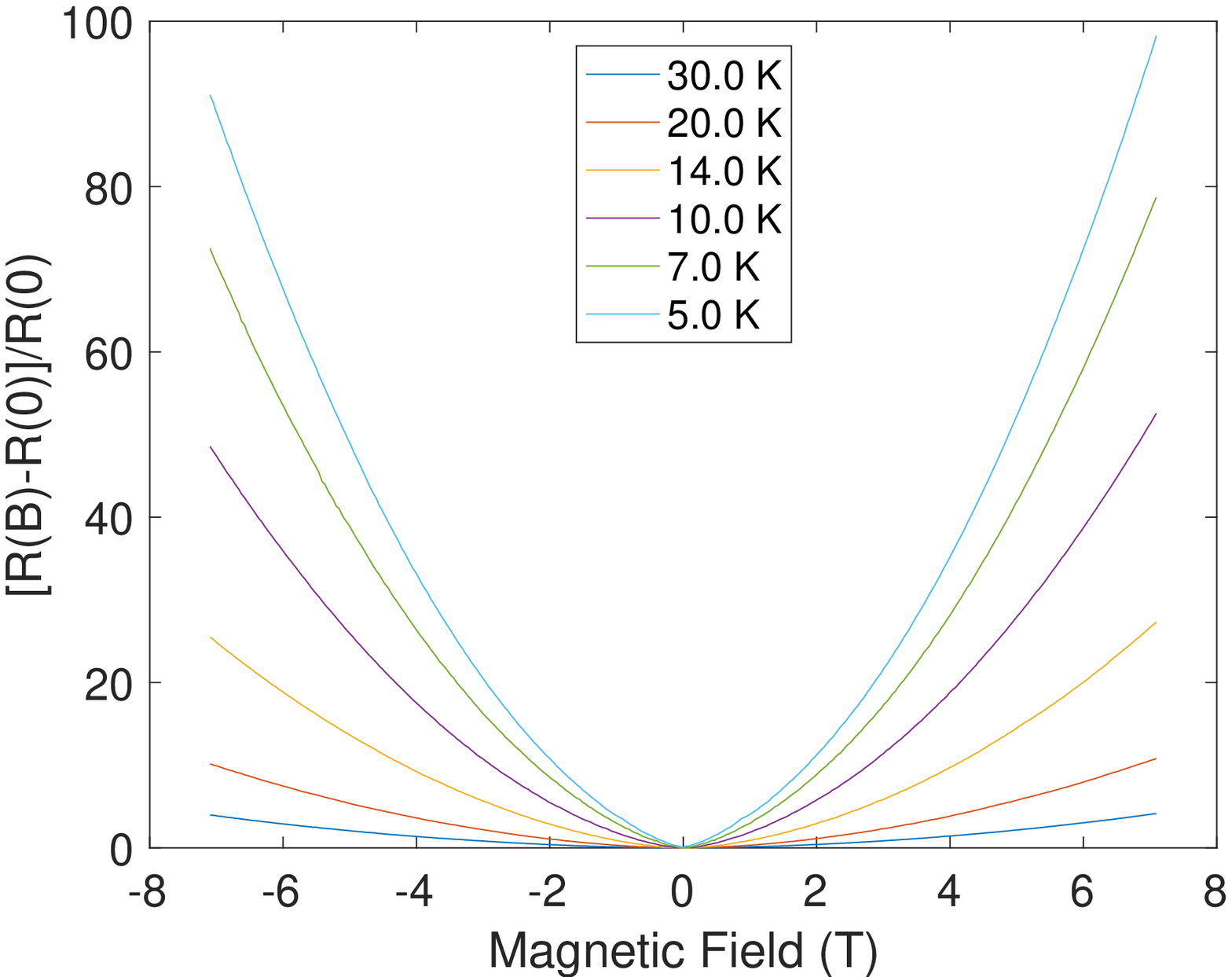}
\caption{Dependence of magnetoresistance on the magnetic field at various temperatures for sample 1.}
\label{mr}
\end{center}
\end{figure}

A typical temperature set of magnetoresistance along the $ab$ in magnetic field perpendicular to this plane is shown in Fig. \ref{mr}. The magnetoresistance is slightly asymmetric as a result of a small asymmetry in the arrangement of contacts and the imperfect shape of the sample. All the data presented below correspond to the magnetoresistance, symmetrized with respect to the direction of the magnetic field. Near $ B = 0 $, the magnetoresistance is quadratic, however, with decreasing temperature, the parabolic region narrows rapidly and there is a distinct tendency to formation of a V-shape, indicating development of a linear contribution that is proportional to the absolute value of the magnetic field.
Also, with decreasing temperature, the temperature evolution of magnetoresistance practically disappears, which is consistent with the results of measurements \cite{MR}. Measurements of 2 and 10 times lower current through the sample showed that the disappearance of the temperature dependence is not associated with the Joule heating of the sample.

\begin{figure} 
\begin{center}
\includegraphics[width=0.45\textwidth]{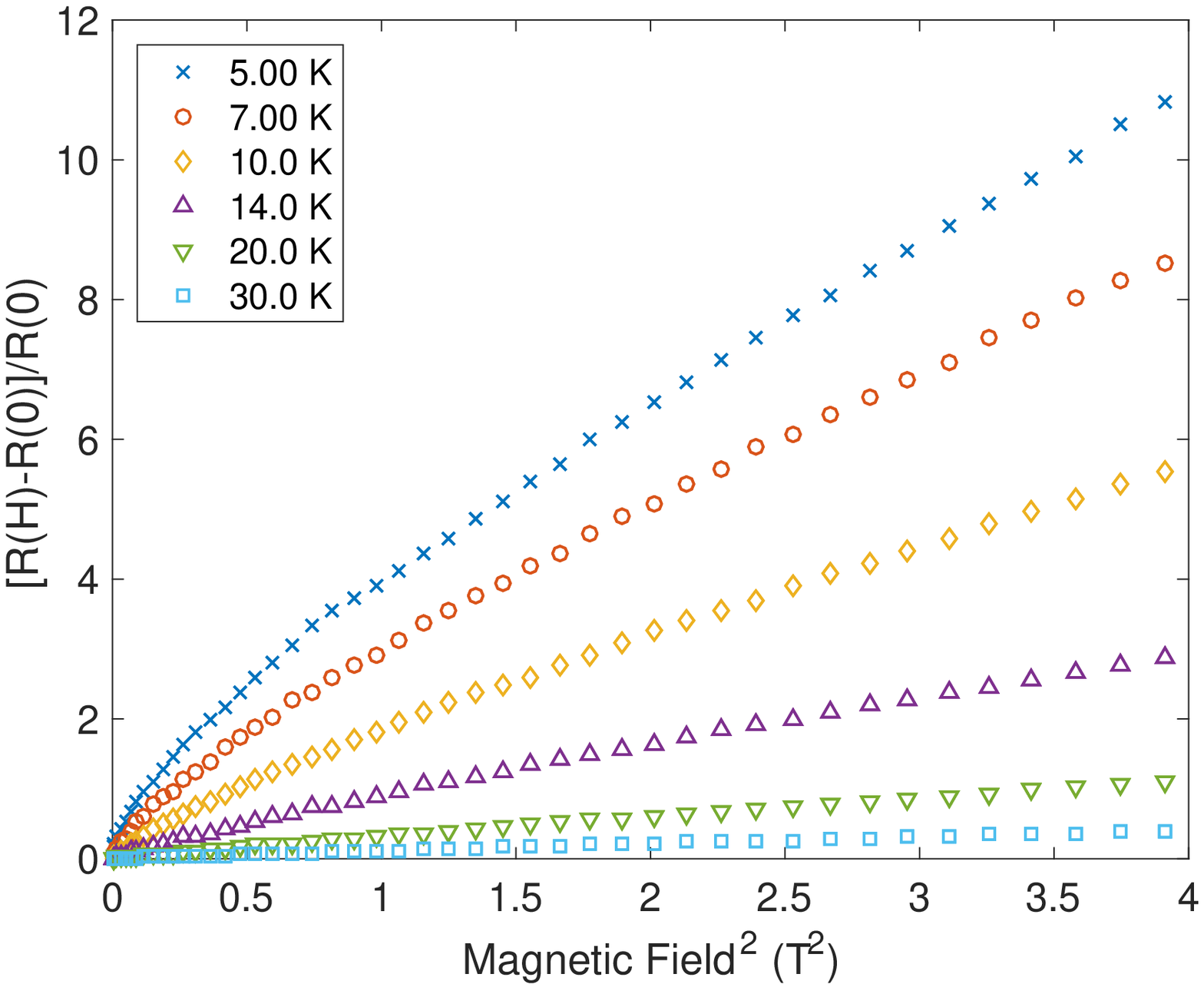}
\caption{Dependence of magnetoresistance on the magnetic field squared at various temperatures. Sample 1.}
\label{mrh2}
\end{center}
\end{figure}

The absence of parabolicity in the middle fields is clearly seen in the Fig. \ref{mrh2}, in which the initial section of the same data is presented as a function of $B^2$. On such a scale, the dependence should be linear, but in practice, a significant deviation from linearity, i.e. from the dependence $\Delta R/R \equiv [R(B) -R(0)]/R(0)\propto B^2$ is observed. An increase in the slope in the low-field region indicates a significantly greater curvature of the $\Delta R(B) $ dependence near zero.

\begin{figure} 
\begin{center}
\includegraphics[width=0.45\textwidth]{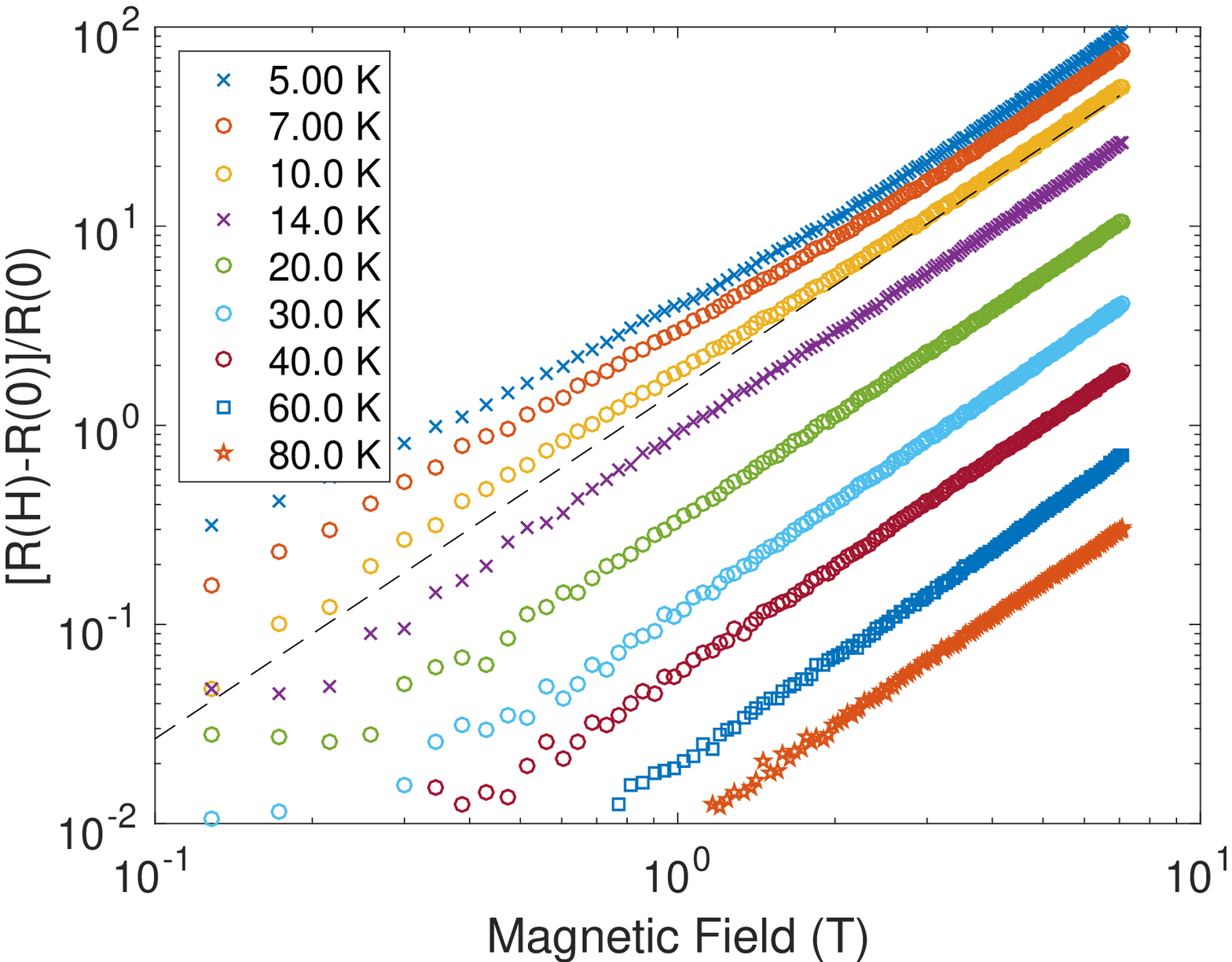}
\caption{Dependences of magnetoresistance on the magnetic field, measured at different temperatures. Sample 1. The dashed line shows the dependence $\Delta R\propto B^{1.75}$.}
\label{mrloglog}
\end{center}
\end{figure}

The temperature set of  the magnetoresistance is plotted in Fig. \ref{mrloglog} in a double logarithmic scale. It can be seen that at all temperatures in the region $ B\gtrsim 2$ T, it is described by the power law $\Delta R\propto B^\alpha$ with the exponent $ \alpha \approx 2$.

The presented results indicate a relatively complex functional dependence of magnetoresistance on the magnetic field. It can be seen that at least for the lowest temperatures, the parabolic section of large curvature in the region of very small magnetic fields $B\lesssim 0.1$ T is replaced by close to linear growth. In turn, in the large fields region $B\gtrsim 1$ T, the linear growth is replaced by a close to parabolic power law, but with less curvature.

\begin{figure} 
\begin{center}
\includegraphics[width=0.45\textwidth]{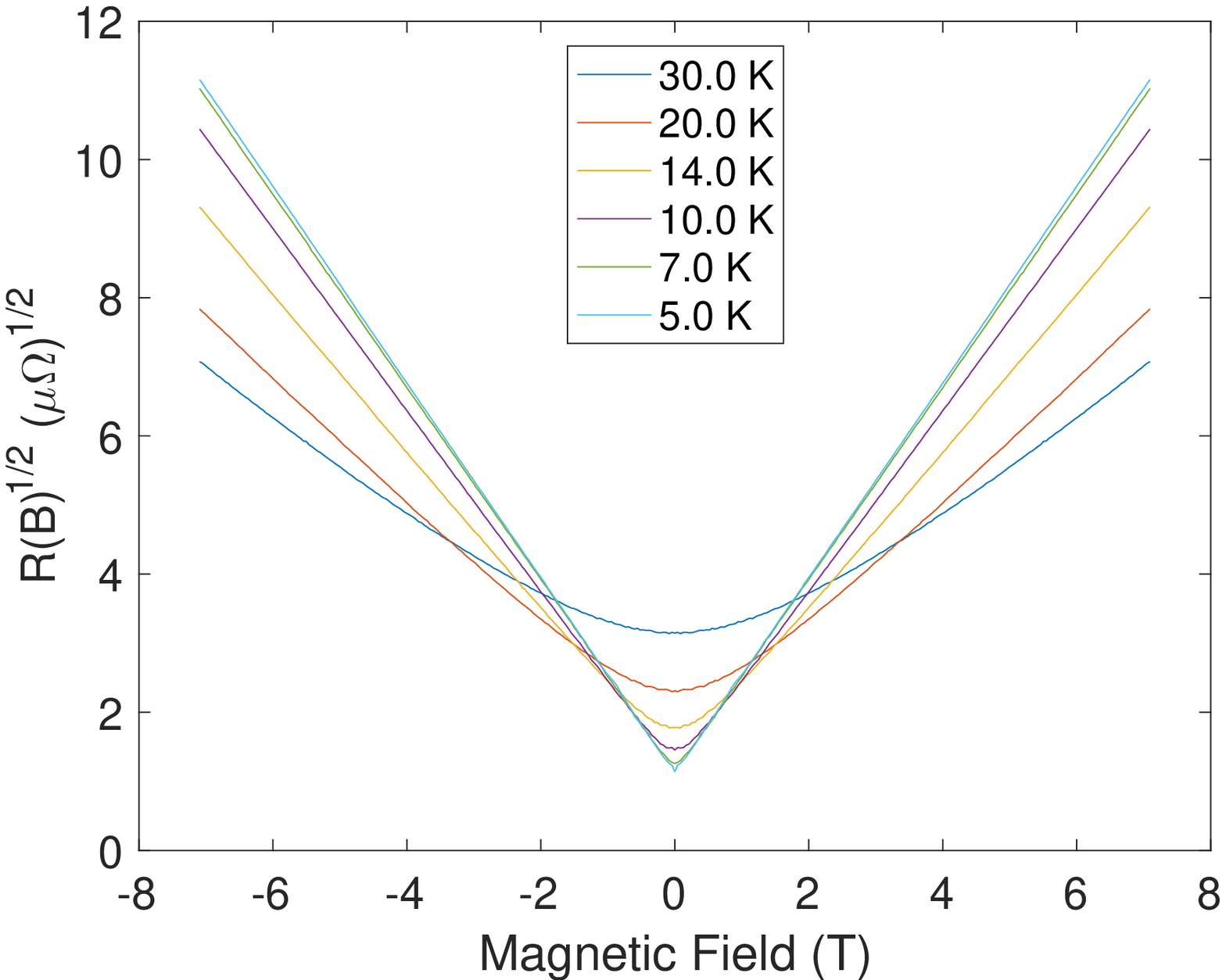}
\caption{Dependences of $ R^{1/2} $ on the magnetic field, measured at various temperatures. Sample 1.}
\label{mrsqrt}
\end{center}
\end{figure}

The dependence of $R(B,T)^{1/2}$ on the magnetic field is shown in Fig. \ref{mrsqrt}. This dependence, at least for the lowest temperatures, has a simple form resembling a hyperbole and consisting of two branches asymptotically approaching linear dependences and connected by a transition region near $ B = 0 $. For the sample 2, the exponent providing the best approximation is slightly higher and corresponds to $ 1/\alpha = 0.55$.

Such a behavior allows us to consider the approximation of the entire dependence using an equation
\begin{equation}
R(T,B)=R_0(T)\left(c(T)+\sqrt{1+\eta ^2B^2}\right)^\alpha,
\label{eq:main}
\end{equation}
which describes the above features of the magnetoresistance. Here $B$ is the magnetic field induction, $c$ is a temperature-dependent dimensionless parameter, $\eta$ is a parameter with the same unit dimensionality as mobility, $ R_0(T)$ is related to the resistance at zero field by the relation $R_0(T) = R(T, 0)/(c + 1)^\alpha$, and $\alpha \approx 2$. For $\alpha = 2$ and $\eta B\gg 1 $, the magnetoresistance described by the equation \ref{eq:main} can be written as
\begin{equation}
\frac{\Delta R}{R} \approx \frac{2\eta c}{(c+1)^2}B + \frac{\eta^2}{(c+1)^2} B^2,
\label{eq:lin2}
\end{equation}  
i.e. reduces to the form $\Delta R/R = aB + bB^2 $ containing a term linear in the magnetic field. Such a description of magnetoresistance was proposed for Dirac semimetals with the nodal line ZrSiSe and ZrSiTe \cite{Ali}, and also theoretically obtained for this class of materials \cite{LMR}.
For $c = 0$ and $\alpha =2$,  the equation \ref{eq:main} describes the quadratic transverse magnetoresistance expected in topologically trivial semimetals with equal electron and hole concentrations (see, for example, \cite{MR}). Thus, the equation \ref{eq:main} is a phenomenological extension of the equation used to describe the magnetoresistance of topologically trivial materials. In its turn, for $\alpha = 2$, the equation \ref{eq:main} is a particular case of the more general equation  
\begin{equation}
R(B)=R_0+R_1\sqrt{1+\eta_1^2B^2}+bB^2, 
\label{eq:general}
\end{equation}
in which the zero-field resistance is determined by the sum $R_0 + R_1$, the second term describes both the linear contribution and the transition region in the vicinity of $B = 0$, and the third one describes the usual quadratic magnetoresistance.

\begin{figure} 
\begin{center}
\includegraphics[width=0.45\textwidth]{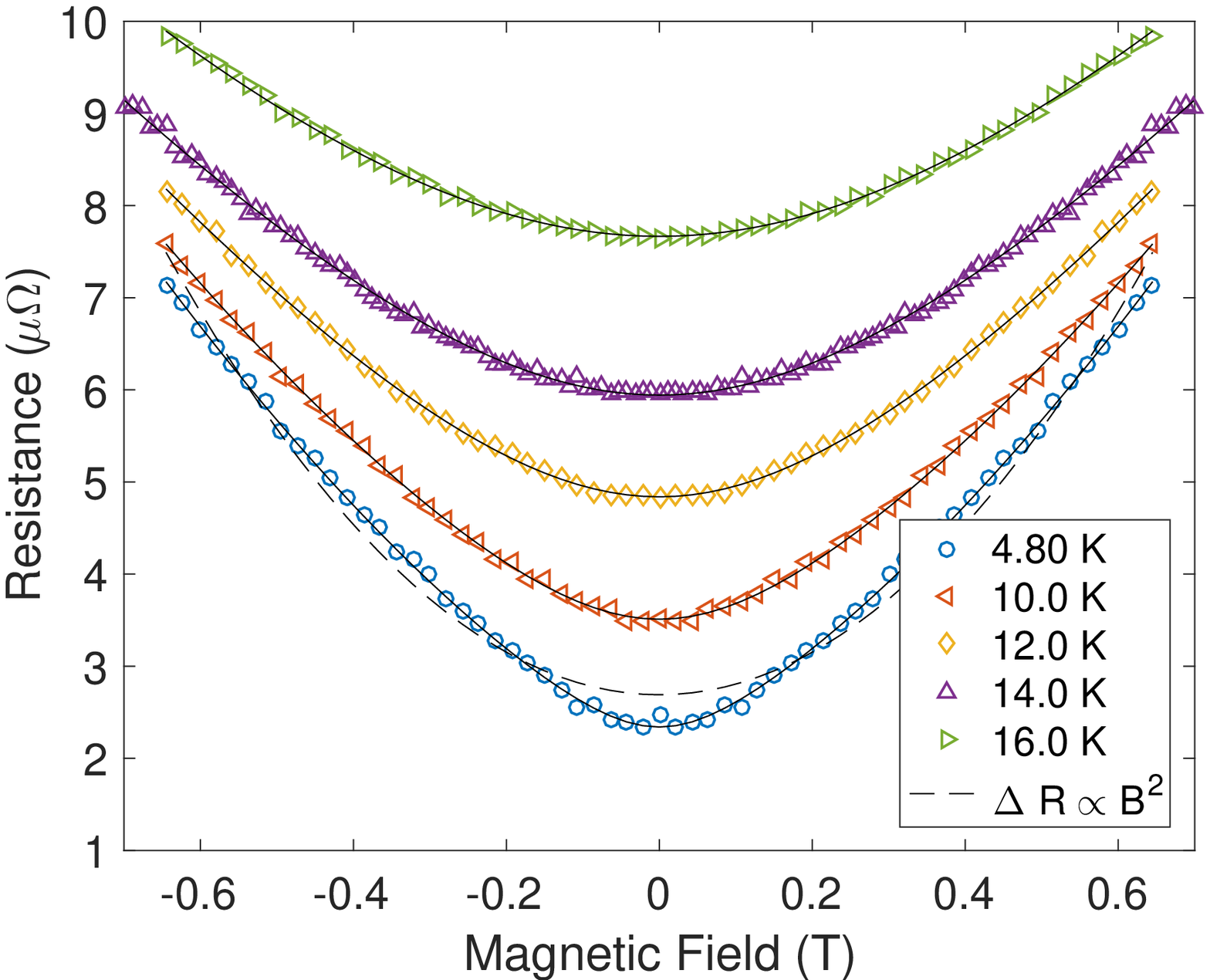}
\caption{Magnetoresistance in the region of weak magnetic fields. The solid lines show the approximation using the equation \ref{eq:main}. The dotted line for comparison shows the quadratic approximation of the data obtained at $ T=4.8$~K. Sample 2.}
\label{smallmr}
\end{center}
\end{figure}

It turns out that the equation \ref{eq:main} describes all the available data with a relative error of the order of a few \% \footnote{The largest deviation is due to corrections arising below the temperature of the superconducting transition of the low-melting-temperature solder used for soldering the contacts.}. On a scale of Fig. \ref{mr},\ref{mrsqrt}, the approximation is practically indistinguishable from the measurement results. Fig. \ref{smallmr} shows the low-field data and their approximation. It can be seen that the approximation accuracy is determined by the noise level of the resistance measurement (in this case, about 50 n$\Omega$, which is 1\% of the characteristic resistance of 5 $\mu \Omega$). The values of the approximation parameters at $T=4.8$~K are $c=6.4$, $\eta =10^5$~cm$^2$/V$\cdot$s. With increasing temperature, both values decrease.

In topologically trivial compensated semimetals, the curvature of the dependence $\Delta R(B)$ is constant and determined by mobility, as $\Delta R(B)/R(0)=\mu_e\mu_h B^2$ (see for example \cite{MR}).
For $\eta B\ll 1$, the equation \ref{eq:main} leads to $\Delta R/R=\eta^2 B^2 /(c + 1)$, which corresponds to the case of quadratic magnetoresistance with mobility $\mu_0 =\eta /\sqrt {c + 1} $. For $\eta B \gg 1$, $\mu_\infty =\eta /(c+1)$. Since $c>  0$, then $\mu_0 =\mu_\infty\sqrt{c + 1}$. Such a change in the curvature of the dependence $\Delta R(B)$ by several times corresponds to a change in the slope of the curves in Fig. \ref{mrh2}. Regions of a higher zero-field curvature can be seen in many topologically non-trivial materials, such as Bi \cite{bi}, NbP \cite{nbp}, MoTe$_2 $ \cite{mote2}, Cd$_2$As$_2$ \cite{cdas}, mentioned above ZrSiSe and ZrSiTe \cite{Ali}, InBi \cite{MR} and many other materials. For this reason, it can be assumed that the equations \ref{eq:main},\ref{eq:general} describes the magnetoresistance of a wide range of topologically non-trivial materials.

The results obtained in this work indicate the realization of a more complex dependence of the transport properties of the node-line Dirac semi-metal InBi on the magnetic field than the quadratic one. The observed deviation from the quadratic dependence corresponds to a linear contribution, which is expected for the nodal-line Dirac systems \cite{LMR}. It is the presence of such a contribution that explains the non-parabolicity of the magnetoresistance in magnetic fields of the order of 0.1-1 Tesla (see also the equation \ref{eq:lin2}). The description of magnetoresistance using the equations \ref{eq:main} proposed in this paper does not currently have a theoretical basis. Nevertheless, for $c=0$ this equation describes the quadratic magnetoresistance of topologically trivial materials, and for $c\neq 0$ it allows us to describe with practical accuracy all the detected features of the magnetoresistance of the  Dirac node-line semimetal InBi.

The authors are grateful to A.A. Mayzlakh for help in crystal growth, V.A. Luzanov for X-ray analysis of the crystals, V.V. Pavlovsky, N.I. Fedotov and A.A. Mayzlakh for useful discussions. This work was supported by the Russian Science Foundation (grant 16-12-10335).

\end{document}